\documentclass{elsart}
\usepackage{graphicx,amssymb}
\newcommand{\be}{\begin{equation}}
\newcommand{\ee}{\end{equation}}
\newcommand{\bea}{\begin{eqnarray}}
\newcommand{\eea}{\end{eqnarray}}
\newcommand{\comment}[1]{}
\newcommand{\fra}[2]{{\textstyle\frac{#1}{#2}}}

\begin{document}
\begin{frontmatter}

\title{From the conserved Kuramoto-Sivashinsky equation to a
coalescing particles model}
\author{Paolo Politi}
\ead{paolo.politi@isc.cnr.it}
\address{Istituto dei Sistemi Complessi, Consiglio Nazionale
delle Ricerche, Via Madonna del Piano 10, 50019 Sesto Fiorentino, Italy}
\author{Daniel ben-Avraham}
\ead{benavraham@clarkson.edu}
\address{Physics Department, Clarkson University, Potsdam, NY 13699-5820}

\begin{abstract}
The conserved Kuramoto-Sivashinsky  (CKS) equation,
$\partial_t u = -\partial_{xx}(u+u_{xx}+u_x^2)$, 
has recently been derived in the context of crystal growth,
and it is also strictly related to a similar equation appearing,
e.g., in sand-ripple dynamics.
We show that this equation can be mapped into the motion of a
system of particles with attractive interactions, decaying
as the inverse of their distance.
Particles represent vanishing regions
of diverging curvature, joined by arcs of a single parabola, 
and coalesce upon encounter.
The coalescing particles model is easier to simulate than the
original CKS equation.  The growing interparticle distance $\ell$ 
represents coarsening of the system, and we are able to establish firmly
the scaling $\bar\ell(t) \sim \sqrt{t}$.
We obtain its probability distribution function, $g(\ell)$, numerically,
and study it analytically within the hypothesis of
uncorrelated intervals, finding an overestimate at large distances.
Finally, we introduce a method based on coalescence waves which
might be useful to gain better analytical insights into the model.
\end{abstract}

% 02.50.Ey	Stochastic processes
% 05.45.-a 	Nonlinear dynamics and chaos
% 05.70.Ln	Nonequilibrium and irreversible thermodynamics
% 81.10.Aj	Theory and models of crystal growth; physics of crystal growth,
%		crystal morphology, and orientation

\end{frontmatter}

% main text

\section{Introduction}
\label{sec_intro}

The study of growth processes of crystal
surfaces~\cite{review,JCG_OPL,Watson} has turned out to be a source of
a variety of nonlinear dynamics.
A first, general distinction should be made between a crystal growing
along a high symmetry orientation (e.g., the face (100) of iron)
and one growing along a vicinal orientation (e.g., the face (119) of
copper). In the former case, growth proceeds~\cite{Evans,MK} layer-by-layer via
nucleation, aggregation of diffusing adatoms and coalescence of
islands;
in the latter, the surface is made up of a train of
steps~\cite{Williams} which advance through the capture of diffusing 
adatoms
(step-flow growth).

The interest in the nonlinear dynamics of a crystal surface mainly
comes from the observation that growth is often unstable~\cite{libro_JV}.
Step-flow growth plays a special role because it 
allows for rigorous treatments
and the original two-dimen\-sio\-nal character of the growth 
may reduce to effective one-dimensional equations:
an equation for the density of steps when steps keep straight
and an equation for the step profile, when steps move in phase.
The two cases occur during step bunching~\cite{libro_JV} 
and step meandering~\cite{JPC,BZ}, respectively. 

As for the nonlinear dynamics resulting from the instabilities,
they may vary from spatio-temporal chaos~\cite{BMV} 
to the formation of stable structures~\cite{Uwaha},
from coarsening processes~\cite{coarsening} due to phase instabilities~\cite{PRL}
to diverging amplitude structures~\cite{amplitude}.
Recently, T.~Frisch and A.~Verga have found~\cite{FV,FV2} that in special 
limits\footnote{In the case of vanishing desorption and weak asymmetry
in the attachment kinetics to the steps.} the profile $u(x,t)$ 
of the wandering steps satisfies the equation
\be
\partial_t u = -\partial_{xx} ( u+u_{xx}+u_x^2 ) ~,
\label{cks}
\ee
now known as the {\it conserved Kuramoto-Sivashinsky} (CKS)
equation. The ``conserved" label becomes clear upon comparison with the standard Kuramoto-Sivashinsky eq.,
$
\partial_t u = -\partial_{xx} ( u+u_{xx}) +u_x^2
$.

An equation similar to (\ref{cks}), with an
extra propagative term $\gamma u_{xxx}$, arises in step bunching
dynamics with vanishing desorption~\cite{Misbah1} and in the
completely different domain of sand-ripple dynamics~\cite{Misbah2}. The
propagative term can be removed using the transformation
$u\to u+ (\gamma/2)x$, which, however, introduces $\gamma-$dependent
boundary conditions~\cite{Misbah1}. 
Numerics and heuristic/similarity arguments give
a coarsening pattern whose typical length scale grows
as $\bar\ell \sim t^n$, with a coarsening exponent $n={1\over 2}$,
both in the presence~\cite{Misbah1,Misbah2} and in the absence~\cite{FV} 
of the propagative term.

The linear stability spectrum of the CKS eq. has the form
$\omega= q^2 -q^4$. In many equations having the same $\omega (q)$,
coarsening occurs because the
branch of steady states has a wavelength which is
an increasing function of the amplitude. These steady states
are unstable with respect to phase fluctuations and the profile
evolves in time, keeping close to the stationary branch~\cite{PRE}.

For the above reason, our first step will be to discuss the periodic 
stationary states
of the CKS eq. (Section \ref{sec_ss}), showing they look like sequences
of arcs of a universal parabola, connected by regions of diverging
curvature (asymptotically, angular points).
Direct simulations of the CKS eq.~\cite{FV} show that: 
(a)~during the dynamical evolution, the interface profile can be thought
of as a superposition of parabolas, and (b)~the typical size $\ell$ of
parabolas (in the $x$ direction) grows with time as
$\ell(t) \simeq \sqrt{t}$.

In order to understand better the dynamics, we have simplified
the problem, starting from the observation that angular points can be
seen as effective particles interacting through the connecting
arcs of the parabolas. 
The correspondence of a Partial Differential Equation (PDE) with a system
of particles is known for the deterministic Kardar-Parisi-Zhang equation
\cite{Medina}, where arcs of parabolas are separated by cusps. In that case,
the PDE is linearly stable, which translates to parabolas of decreasing
curvature. For the Burgers equation~\cite{Burgers}, cusps are replaced by
shock waves and arcs of parabolas by linear pieces of decreasing slope.

In principle, the particles
move in two dimensions (the
$\hat{xu}$ plane) and each particle interacts with all other particles.
This full description, if possible, would be exact.
However, we keep things simpler by limiting to the horizontal motion
and to nearest neighbour interactions. Therefore, in Section~\ref{sec_pm}
we show that the dynamics of particles is described by the
equations $\dot x_i = (x_{i+1}-x_i)^{-1} - (x_i - x_{i-1})^{-1}$,
where $x_i$ is the coordinate of the $i-$th particle and $x_{i+1}\ge x_i$.
When two particles collide ($x_i=x_{i+1}$), 
they coalesce and the total number of particles
decreases by one, thus leading to a coarsening process.

Subsequently, we simulate the particle model (Section \ref{sec_sim}) 
finding the coarsening law $\bar\ell(t)$ and the size distribution of
interparticle distances, $g(\ell/\bar\ell(t))$. A very crude approximation
of the Fokker-Planck equation for the particle system (Section \ref{sec_FP})
gives the correct expression for $\bar\ell(t)$, but overestimates $g(s)$
at large $s$. Finally, we introduce the method of coalescence waves
(Section \ref{sec_cw}), finding some preliminary numerical results,
which might guide a future, more rigorous analytical study.

\section{Steady states}
\label{sec_ss}

The steady states of the CKS equation~(\ref{cks}) satisfy the
second-order nonlinear differential equation
$
 u+u_{xx}+u_x^2 = a + b x ~.
$
Since the constant $b$ must vanish in order to get bounded solutions,
while the constant $a$ can be trivially absorbed into a uniform shift
of $u(x)$ and be set to zero, the problem reduces to solving the
differential equation
\be
 u_{xx} = -u - u_x^2 ~.
\label{newton}
\ee
This equation also gives the steady states of a different PDE, 
studied by Mikishev and Sivashinsky~\cite{Mikishev}.
% (Eq.~(\ref{cks}) without the $-\partial_{xx}$ term on the right).
Therefore, we limit ourselves to just a few results that play a major role in what is to follow.

Interpreting $x$ in Eq.~(\ref{newton})
 as  {\it time\/}, the equation corresponds to a
harmonic oscillator subject to an external force proportional to
the velocity squared. Deriving~(\ref{newton})
with respect to $x$ and putting $u_{xx}\equiv{F}$, one obtains
\be
 \frac{du}{dF}= -\,\frac1{1+2F}~.
\ee
Assuming the initial conditions $u(0)=A$ and $u_x(0)=0$, 
the solution is
\be
 F = -u-u_x^2 = -\fra12 + \left(\fra12-A\right)~ e^{2(A-u)} ~.
\label{force}
\ee

We then have the trajectories in the $(u,u_x)$-phase space:
\be
 u_x^2 = \fra12 -u - \big(\fra12-A\big)\,e^{2(A -u)} ~.
\label{qp}
\ee

\begin{figure}
\includegraphics*[width=12cm]{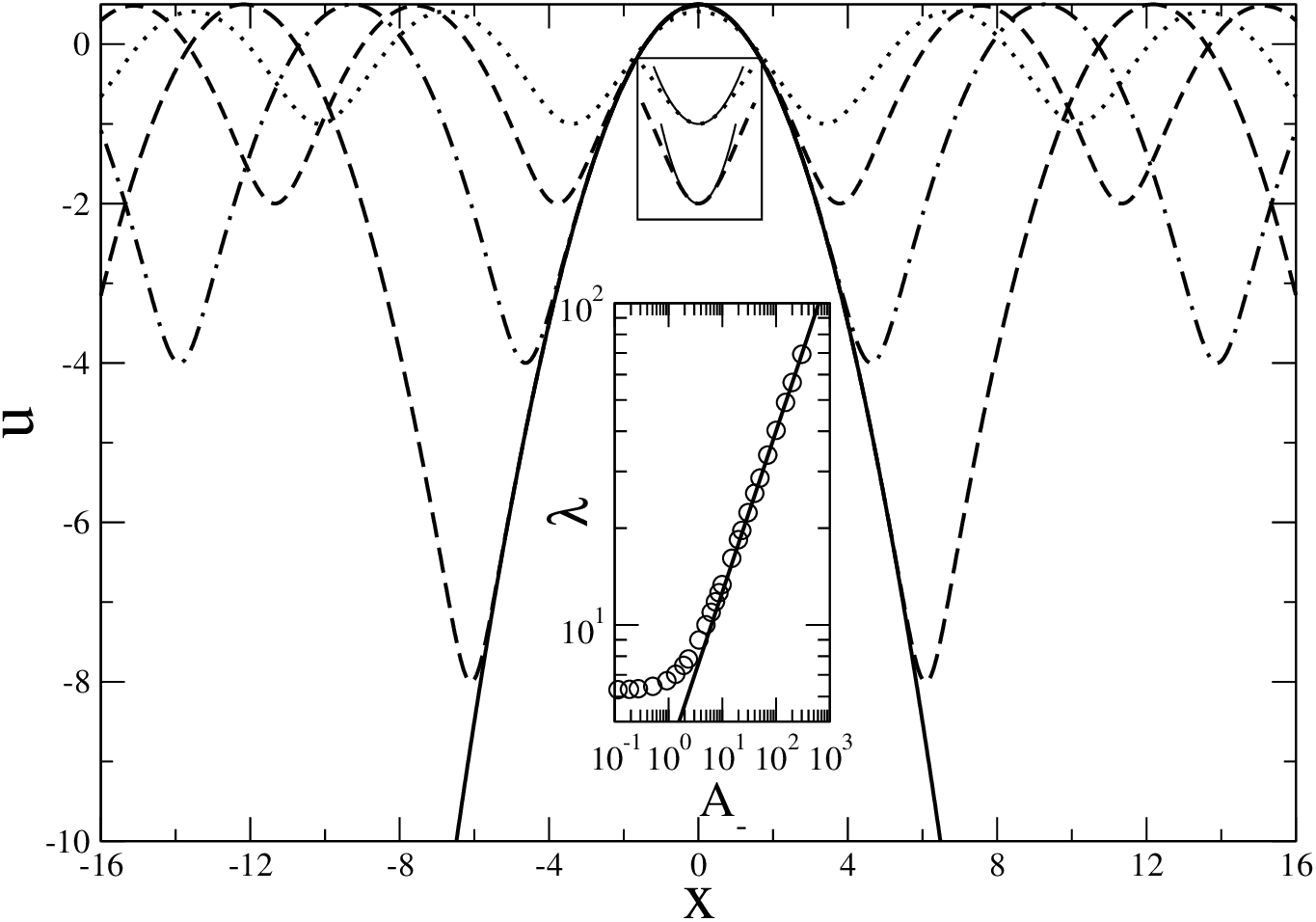}
\caption{Periodic steady states configurations, corresponding to
$A_-=1$ (dotted), $A_-=2$ (short dashed), $A_-=4$ (dot dashed), and
$A_-=8$ (long dashed). The thick full line corresponds to the
limiting parabola $u={1\over 2} -x^2/4$. Small upper inset: the minima of
the configurations and the approximations $u=-A_- + A_-x^2/2$ (thin
full lines). Large lower inset: wavelength $\lambda$ as a function of the
amplitude $A_-$. The full line is the asymptotic relation
$\lambda=4\sqrt{A_-}$. }
\label{fig_traiettorie}
\end{figure}

If $A\ge\fra12$ the force $F$ is strictly
negative and the trajectory is not limited. On the other hand, if
$A=A_+$, with $0<A_+<\fra12$\,, we get periodic, 
bounded trajectories,
which oscillate between $A_+$ and $-A_-$.
For $A_+=\fra{1}{2}$, we obtain the separatrix
$u_x^2=\fra12-u$,
which corresponds to the parabolic trajectory
$u(x)=\fra12-(x-x_0)^2/4$. See Fig.~\ref{fig_traiettorie} for more details.

It is useful to determine the amplitude
$A_-$ in the negative $x$ direction, as a function of $A_+$.
For $A_+\to 0$, it is easily
found that $A_- \to A_+$, while in the important limit
$A_+\to\fra{1}{2}$, $A_-$ diverges logarithmically according to
$A_-e^{-2A_-} \simeq \fra12-A_+$, i.e.,
$
 A_- \approx -\fra{1}{2}\ln(1-2A_+) ~.
$

In proximity of $u=-A_-$, bounded trajectories have a minimum with a
curvature that diverges as $A_-\to\infty$, as shown by the expansion
$u=-A_-+\varepsilon$ in Eq.~(\ref{qp}), which gives
$
 u \approx -A_- + {A_-\over 2} (\delta x)^2
 + O\big[A_-^2 (\delta x)^4\big] ~.
$
This approximation is shown as thin full lines in
Fig.~\ref{fig_traiettorie} (small upper inset). The
quadratic and quartic terms are of the same order when
$|\delta{x}|\approx{1}/\sqrt{A_-}$, which sets the size of the
high-curvature region. In fact, the slope
$u_x(\delta{x}=1/\sqrt{A_-})\approx\sqrt{A_-}$ joins the corresponding
slope of the limiting parabola $u=\fra{1}{2}-x^2/4$, when
$u\approx-A_-$.

Finally, in the large lower inset of Fig.~\ref{fig_traiettorie} we plot the
wavelength $\lambda$ of the steady states as a function of their
amplitude $A_-$. The full line, $\lambda=4\sqrt{A_-}$, gives the
analytical approximation valid for large $A_-$. It can be determined
from the asymptotic parabola $u=\fra{1}{2}-x^2/4$, imposing
$u(\lambda/2)=-A_-$.

\subsection{Steady states and dynamics}
\label{sec_dynamics}

In the Introduction we have argued that steady states are important
because dynamics proceeds by evolving along the family of steady states
of increasing wavelength $\lambda$.\footnote{This is both an observation
(see the following discussion on Fig.~\protect\ref{fig_arcs}) and a
consequence of the stability of steady states with respect to amplitude 
fluctuations, see Ref.~\protect\cite{PRE}.}
In the case of the CKS equation
special attention should be paid to the constant $a$ 
and to the conserved character of Eq.~(\ref{cks}).
We now show how the conservation law fixes the value of
$a$, as a function of $\lambda$. 
We shall then argue that $a$ is the time-dependent vertical
shifting of the surface profile.

On the one hand, the conserved dynamics requires that the spatial
average $\langle u(x,t) \rangle$ is time independent; on the other
hand, steady states $u(x)$ found in the previous section have a
non vanishing and $\lambda$-dependent average value. 
Therefore, the family $u_d(x)$ of steady states which is relevant to the
dynamics is
$
u_d (x) \equiv u(x) + a(\lambda),
$
where the constant $a$ satisfies the condition
$a(\lambda) = - \langle u(x)\rangle$.
For large $\lambda$, the average value of $u(x)$ can be safely determined by
approximating it with the arc of the (separatrix) parabola, so that
\be
 \langle u(x) \rangle =
 \frac1\lambda \int_{-\lambda/2}^{\lambda/2} dx\,
 \bigg(\frac12 - {x^2\over 4}\bigg) + o(\lambda^2)
 = -{\lambda^2\over 48} + o(\lambda^2)~.
\ee
Therefore, for large $\lambda$ we get $u_d(x) = u(x) + \lambda^2/48$.

\section{The particles model}
\label{sec_pm}

\begin{figure}
\includegraphics*[width=12cm]{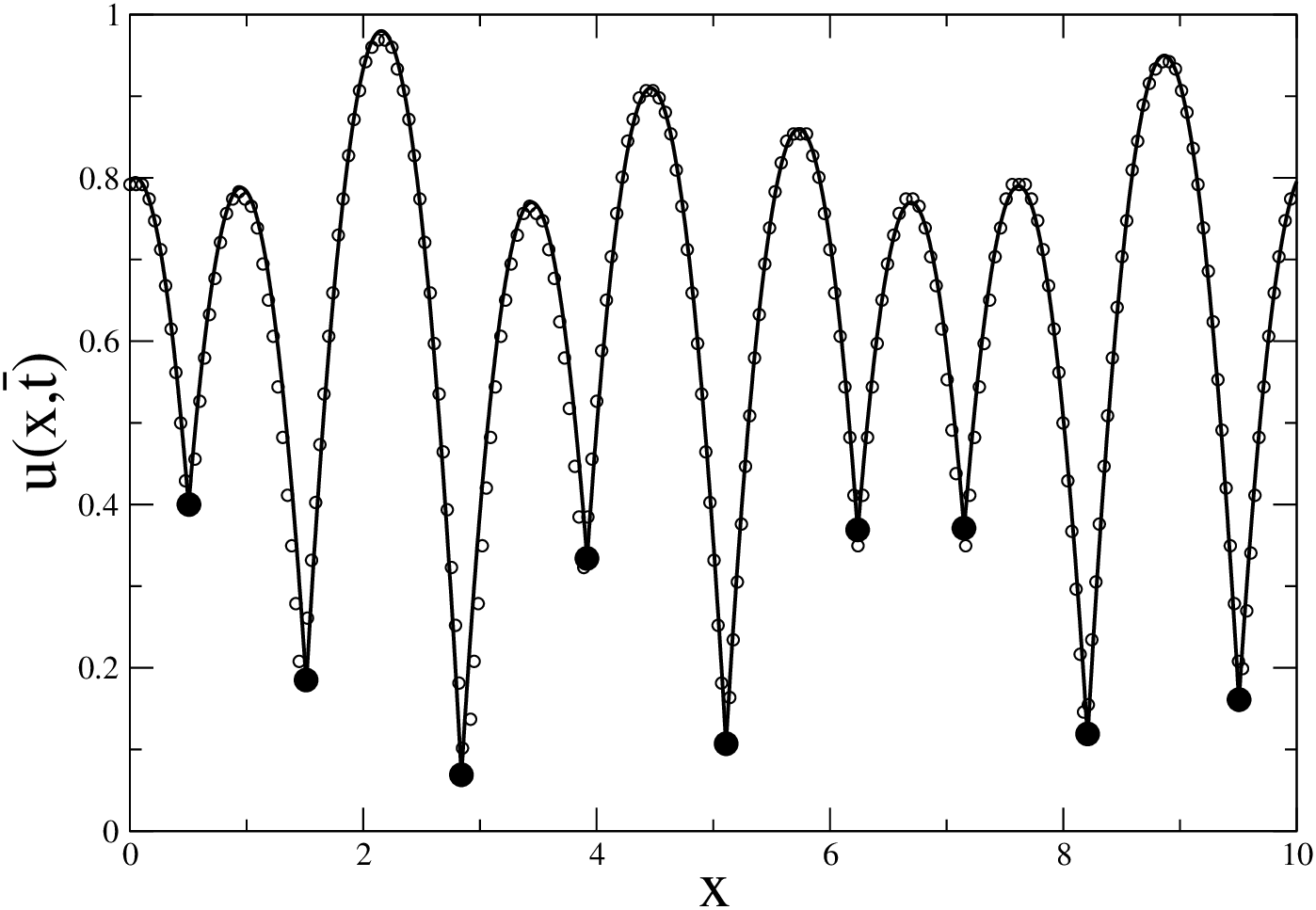}
\caption{Empty circles: The profile emerging from the dynamical
evolution of the CKS eq. has been obtained digitalizing a late time
profile in Fig.~3 of Ref.~\protect\cite{FV}. The full line is a
sequence of arcs of the same parabola ($x$ and $u$ scales are
arbitrary). Big full dots are located at the joins of the arcs and represent
the ``particles".}
\label{fig_arcs}
\end{figure}

The following approach is founded on two observations, the first
based on theory, the second on numerics:
(i) When increasing the wavelength $\lambda$ of steady states,
$u(x)$ tends to a sequence of arcs of the universal parabola
$u(x) = a - (x-\bar x)^2/4$, connected by regions of diverging positive
curvature whose size $\delta \approx 1/\lambda$ is vanishing small.
See Section~\ref{sec_intro} and Fig.~\ref{fig_traiettorie}.
(ii) Dynamics deforms the above picture, but interface profiles
can still be thought of as a sequence of points $(x_n,y_n)$, with
$x_{n+1}> x_n$ and the following properties: 
$u''(x_n)\to\infty$ with increasing time, and between any pair $x_n,
x_{n+1}$ of consecutive points $u(x)$ can be approximated as 
an arc of the universal parabola,
with $a$ and $\bar x$ determined by the conditions $u(x_n)=y_n$ and
$u(x_{n+1})=y_{n+1}$.
See Fig.~3 of Ref.~\cite{FV} and our Fig.~\ref{fig_arcs}.

Therefore points ($x_n,y_n$) (henceforth particles, 
full dots in Fig.~\ref{fig_arcs}) define unambigosuly
the full interface profile and their dynamics should be derivable from
the CKS eq. The coarsening process occurs because bigger parabolas
eat neighboring smaller ones. When one parabola disappears, two particles merge
into one: it is a coalescence process. Because of the conservation
of the order parameter, $d\langle u\rangle/dt=0$, particles do not
move independently and their effective interaction is expected
to be fairly complicated and long-range. 

\begin{figure}
\includegraphics*[width=12cm]{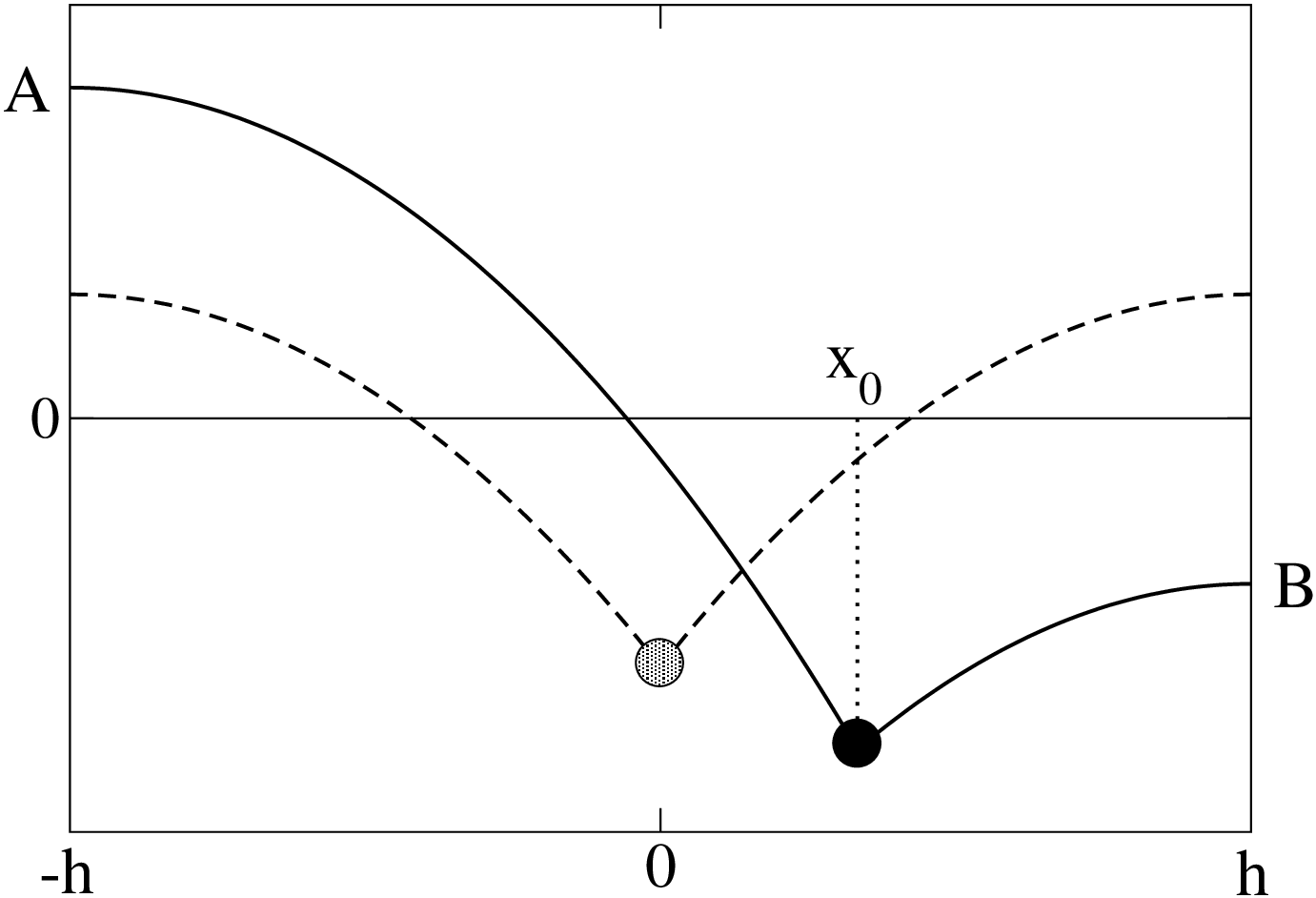}
\caption{Two arcs of the universal parabola, $u_1(x)=A -{1\over 4}
(x+h)^2$ and $u_2(x)=B - {1\over 4}(x-h)^2$, join in $x=x_0$. 
This condition and the conservation of the average value
$\langle u(x)\rangle$ fix two of the three parameters
$A,B,x_0$. The dashed line corresponds to $x_0=0$, the symmetric (and unstable)
configuration.}
\label{fig_two_arcs}
\end{figure}

Here we limit ourselves
to a simple model where a single bigger parabola ($u_1(x)=
A-(x+h)^2/4$) eats a smaller one ($u_2=B-(x-h)^2/4$),
see Fig.~\ref{fig_two_arcs} (full line). The full configuration
is defined by the three parameters $A,B$ and $x_0$, the point where
the two parabolas meet. Two conditions must be fulfilled: continuity
in $x_0$ implies $B=A-hx_0$ and conservation $d\langle u\rangle/dt=0$
implies $A+B-{h^2\over 6}  - {x_0^2\over 2}=0$.
In this picture, the particle in $x=x_0$ has neighbouring
particles at distances $\ell_-=2(h+x_0)$ on the left and
$\ell_+=2(h-x_0)$ on the right.

The next step is to use the CKS eq. to evaluate $d\langle u^2\rangle/dt$:
it is enough to multiply both sides of Eq.~(\ref{cks}) by $u$ and to
integrate. We get
\be
{1\over 2}{d\langle u^2\rangle\over dt} = \langle u_x^2\rangle -
\langle u_{xx}^2\rangle - {1\over 3}\langle (u_x^3)_x\rangle ~,
\label{eq_x0}
\ee
where the last term on the right vanishes because of periodic boundary conditions.
The average values $\langle u^2\rangle, \langle u_x^2\rangle$ and
$\langle u_{xx}^2\rangle$ are evaluated using the two-arcs approximation
depicted in Fig.~\ref{fig_two_arcs} and are therefore functions
of $x_0$ only. Once we replace $(d/dt)$ by $\dot x_0 (d/dx_0)$, we obtain
a differential equation for the position $x_0$ of the particle,
\be
\dot x_0 = {2(\langle u_x^2\rangle - \langle u_{xx}^2\rangle)\over
\displaystyle{d\langle u^2\rangle \over dx_0} } .
\label{vel_x0}
\ee

The evaluation of $\langle u^2\rangle$ and $\langle u_x^2\rangle$ 
is straightforward, because the region of the angular point 
can be safely neglected:
\bea
\langle u^2\rangle &=& {h^2 x_0^2\over 6} - {x_0^4\over 12} 
+ \hbox{function}(h)~, \\
\langle u_x^2\rangle &=& {h^2\over 12} + {x_0^2\over 4}.
\eea

As for the average value $\langle u_{xx}^2\rangle$, we may write 
$
\langle u_{xx}^2\rangle = {1\over 4} + 
\langle u_{xx}^2\rangle_{\hbox{\tiny ang}}
= {1\over 4} + {\delta\over 2h}\left( {\Delta u_x\over\delta}\right)^2 ,
$
where $\langle\cdots\rangle_{\hbox{\tiny ang}}$ means the average on the
angular point and $\Delta u_x = (u_x)_+ - (u_x)_-$ is the abrupt change
of slope occurring through the angular point, on a distance of order
$\delta$. Since $\Delta u_x = u_2'(x_0)-u_1'(x_0)=
-(x_0-h)/2 + (x_0+h)/2=h$, we get
$
\langle u_{xx}^2\rangle = {1\over 4} + {h\over 2\delta} \simeq
{h\over 2\delta},
$
the term $\fra14$ being negligible for large $\lambda$ (which means
large $h$ and small $\delta$).

For a periodic configuration of wavelength $\lambda$, we have
$\delta \simeq 1/\lambda$. In the configuration of Fig.~\ref{fig_two_arcs}
we do not have a single $\lambda$, but the two quantities $\ell_{\pm}$.
If we assume
$\delta \simeq \left( \ell_{+}^{-1} + \ell_{-}^{-1}\right)$,
we get $\delta = \delta_0 h/(h^2-x_0^2)$, with $\delta_0$ 
being determined by the condition that the right-hand-side
of Eq.~(\ref{eq_x0}) vanishes for $x_0=0$: in fact, the particle
has zero speed in the symmetric configuration $x_0=0$.
So, we get
$
\langle u_x^2\rangle - \langle u_{xx}^2\rangle = {h^2\over 12} +
{x_0^2\over 4} + {h^2-x_0^2\over 2\delta_0} = {x_0^2\over 6}.
$
It is worth stressing that a completely different assumption,
$\delta \simeq \sqrt{\ell_{+}^{-1} \ell_{-}^{-1}}$,
gives a result which is almost indistinguishable from this.

We now have to determine the derivative appearing at the denominator
of Eq.~(\ref{vel_x0}): $d\langle u^2\rangle /dx_0
=x_0(h^2-x_0^2)/3$, so we finally get
$\dot x_0 = x_0/(h^2 - x_0^2)$. 
Expressing $x_0$ in terms
of the interparticle distances $\ell_\pm = 2(h\mp x_0)$, we finally get
$\ell_+\ell_- = 4(h^2-x_0^2)$, $(\ell_- - \ell_+)=4x_0$, and
\be
{d x_0\over dt} = {1\over\ell_+} - {1\over\ell_-}.
\ee  

This is one of the main results of the paper: it means that the conserved
Kuramoto-Sivashinsky equation can be translated into the motion
of a system of particles with attractive interactions, decaying
as the inverse of their distance, and undergoing a coalescence
process when they collide. If the coordinate of the $i-$th particle
is $x_i$, we can write
\be
{d x_i\over dt} = {1\over (x_{i+1} - x_i)} - {1\over (x_i - x_{i-1})}.
\label{eq_x}
\ee
Alternatively, if $\ell_i = (x_{i+1}-x_i)$ is the distance between
particles $i$ and $(i+1)$, we get
\be
{d \ell_i\over dt} = {1\over\ell_{i+1}} + {1\over\ell_{i-1}} -
{2\over\ell_i}.
\label{eq_l}
\ee

Since interparticle force decays as the inverse of the distance,
the coarsening law, i.e. the time dependence of the average distance
between particles, $\bar\ell (t)$, can be easily inferred from 
scaling considerations,
\be
{d\bar\ell\over dt} \sim {1\over\bar\ell} ~~~\longrightarrow ~~~~
\bar\ell(t) \sim \sqrt{t}.
\label{ell_scaling}
\ee

This result is in agreement with numerical simulations~\cite{FV} 
of the CKS equation and with numerical results of 
the particle model,
discussed in the following section.
It will also be corroborated by the Fokker-Planck approach, 
Section~\ref{sec_FP}.
It is worth noting that the result $\bar\ell \sim \sqrt{t}$ is {\it not\/}
related to diffusion since the particles motion is strictly deterministic; 
it arises from the $1/\ell$ decay of the interparticle force with distance $\ell$.

\subsection{Simulation of the particle model}
\label{sec_sim}

\begin{figure}
\includegraphics*[width=12cm]{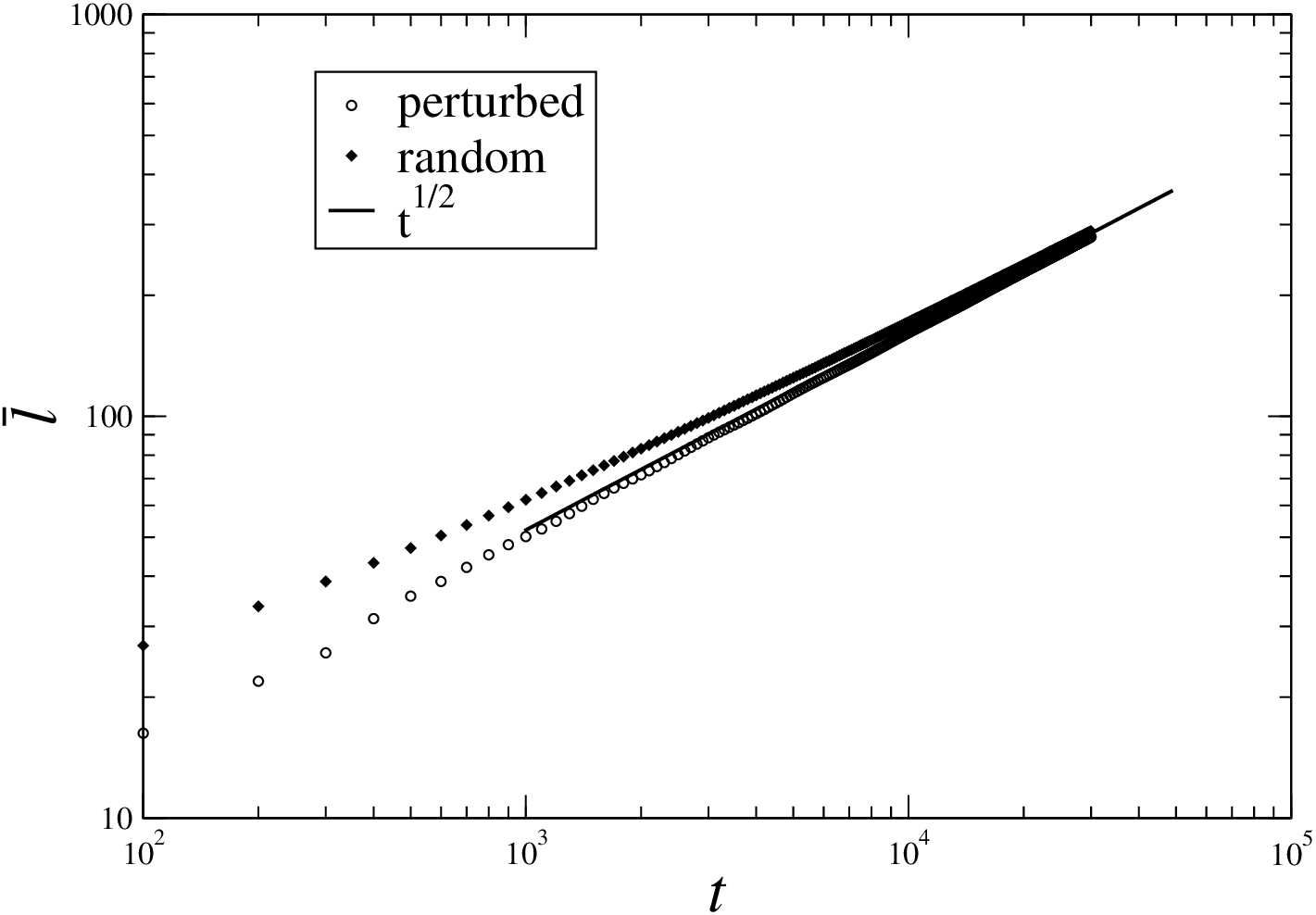}
\caption{
The average distance between particles as a function of time, for
the same initial density but different starting configurations: 
random distribution of particles (full diamonds) and a slightly
perturbed uniform configuration (empty circles). 
The asymptotic law (full line) is $\bar\ell(t)=c_0\sqrt{t}$,
with the same $c_0$ for the two distributions.
}
\label{fig_coarsening}
\end{figure}

Rather than solving Eqs.~(\ref{eq_x}) for particle positions, we have
solved Eqs.~(\ref{eq_l}) for the distances.
Fig.~\ref{fig_coarsening} shows that the expected law
$\bar\ell(t) \sim t^{1/2}$ is satisfied for very different initial
conditions: a random distribution (full diamonds) and a slightly
perturbed uniform distribution (empty circles).
The asymptotic value of $\bar\ell(t)$ is the same for the two
distributions, showing that the prefactor $c_0$ in $\bar\ell(t)=
c_0\sqrt{t}$ only depends on the initial density.
The two sets (circles and diamonds)
are distinct at small $t$ because the initial random distribution favours more
coalescences at short times than the uniform one.

\begin{figure}
\includegraphics*[width=12cm]{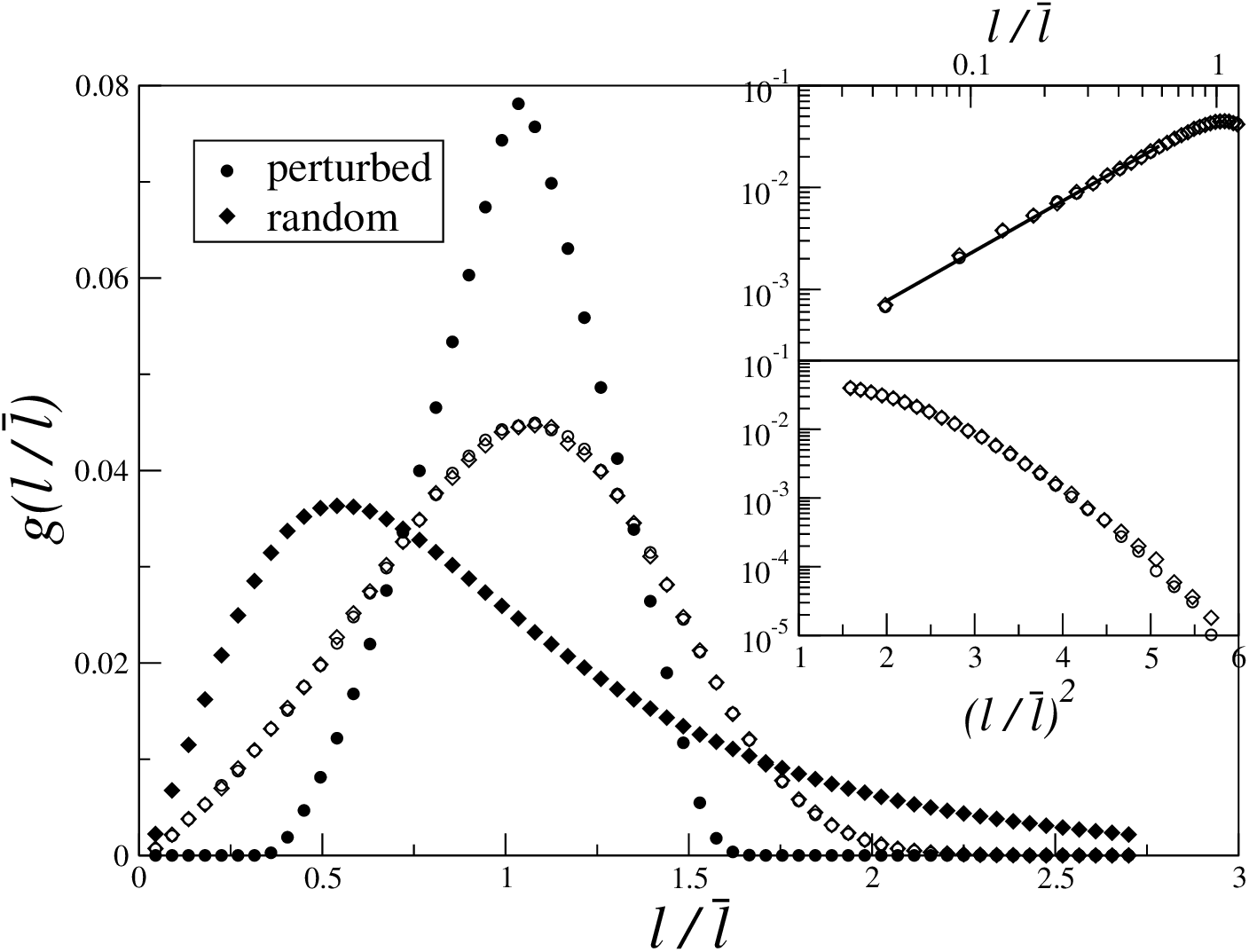}
\caption{
The asymptotic distribution of interparticle distances (empty simbols)
for two different initial distributions (full symbols): 
random (diamonds) distribution and slightly perturbed (circles) 
uniform distribution.  The system length is $L=10^6$ and 
there are $10^5$ particles at the begininng.
The asymptotic distribution is shown at time $t=10^5$, and is averaged 
over hundreds of runs.
Upper inset: $g(\ell/\bar\ell)$ at small distances, on a log-log scale.
Lower inset: $g(\ell/\bar\ell)$ vs.~$(\ell/\bar\ell)^2$ at large distances, on a lin-log scale.
}
\label{fig_distribution}
\end{figure}

Fig.~\ref{fig_distribution} shows the normalized distribution $g(s)$ 
of interparticle distances as a function of $s=\ell/\bar\ell$.
Again, widely different initial configurations produce the same
asymptotic distribution (empty circles and diamonds). As for the
limiting behaviors of $g(s)$, they are plotted in the upper inset for small $s$
and in the lower inset for large $s$. At small $s$ we clearly have a power law
distribution, $g(s) \sim s^\alpha$, with exponent $\alpha\approx 1.3$.
At large $s$, data seem to suggest a gaussian tail.

The qualitative features of the distribution $g(s)$ can be understood by
comparison with the one-species diffusion-limited coalescence process 
on the line~\cite{coal}.  In our case, particles separated by a smaller gap than average
close the gap at increasingly larger speeds.  As a result, the system evolves effectively
as if nearest particle pairs react instantaneously, before the rest of the system evolves (numerical
simulations do confirm this intuitive notion).  In the diffusion-limited coalescence process, 
a quasi-static approximation shows that particle pairs react at a time proportional to the gap
between them.  Thus, both processes favor faster reactions between nearest particle pairs
(in our case, more aggressively so).
The distribution $g(s)$ for diffusion-limited coalescence is known exactly, yielding
$g(s)\sim s$ for small $s$, and a gaussian tail for $s$ large.  It is not surprising that we
find a similar $g(s)$. The faster-than-linear behavior, $g(s)\sim s^\alpha$, in our
case, is consistent with the faster reactions between nearest particle pairs, 
leading to a faster depletion of the probability distribution function near the origin.

\subsection{The Fokker-Planck equation}
\label{sec_FP}

In this Section we use the Fokker-Planck (FP) equation
for the distances $\ell_i$, under the approximation of uncorrelated
intervals. The same approach has already been used for models
of particles interacting with a force decaying exponentially,
see Refs.~\cite{FP,kinks}, and we refer the reader to those
papers for more details.

If $f(\ell)=1/\ell$ is the force between two particles at distance
$\ell$, Eq.~(\ref{eq_l}) can be written as
\be
{d\ell_i\over dt} = f(\ell_{i+1})+f(\ell_{i-1})-2f(\ell_i)
\equiv {\cal U}_i(\{\ell\})
\ee
and the FP equation for the probability $\rho(\{\ell\},t)$
to find a given distribution $\{\ell\}$ at time $t$ is
\be
\partial_t \rho = -\sum_k {\partial\over\partial\ell_k}
\left[ {\cal U}_k(\{\ell\},t)\rho\right] .
\ee

We are mainly interested in the time dependence of the average
distance, $\bar\ell(t)$, and in the probability distribution for
the distances, $g(\ell,t)$, which is expected to have the
asymptotic scaling form
$
g(\ell,t) = n(t)\tilde g\left( \ell/\bar\ell(t)\right)
$,
with $n(t)=1/\bar\ell(t)$. In the approximation of uncorrelated
intervals, we get the equation
$
\partial_t g(\ell,t) = -2{\partial\over\partial\ell}
\left[ \left( f(\bar\ell) - f(\ell) \right) g(\ell,t)\right],
$
which does not include the coalescence process, because the form
$\partial_t g=-\partial_\ell J$ implies the conservation law
$\partial_t\bar\ell=0$.
Details on how to include coalescence are very similar to published
papers~\cite{FP,kinks} for different $f(\ell)$, 
so here we merely state the results:
\bea
\bar\ell(t) &=& \ell_0 t^{1/2}, \\
\tilde g (s) &=& \tilde g_0 {s\over s^2 -c_1 s + c_2}.
\eea

The expression for $\bar\ell(t)$ agrees with numerical results 
(Fig.~\ref{fig_coarsening}) and with scaling considerations, Eq.~(\ref{ell_scaling}).
On the other hand, the results for $g(s)$ do not match our
numerical findings, discussed in Section~\ref{sec_sim},
indicating the importance of correlations, neglected by this approach.
The evolution equation~(\ref{eq_l}) clearly shows that adjacent
intervals are strongly anticorrelated, as gaps grow (or shrink) on 
expense of their surroundings; larger than typical intervals are surrounded by
smaller than typical ones, and vice versa.  Thus, larger than typical intervals
grow far less than the independent interval approximation would allow, as the
neighbors they engulf are typically smaller than average.  In this way
the approximation overestimates the frequency of long intervals ($g(s)\sim1/s$,
instead of a gaussian tail).  Similarly, being surrounded by large
intervals, short intervals contract faster than if they were surrounded by typical
intervals, explaining the overestimate of their frequency ($g(s)\sim s$ instead of $s^\alpha$)
by the approximation.

\subsection{Coalescence waves}
\label{sec_cw}

This final section is dedicated to preliminary results on the study
of coalescence waves, which might be useful for a deeper 
understanding of numerical results for the particles model.

\begin{figure}
\includegraphics*[width=12cm]{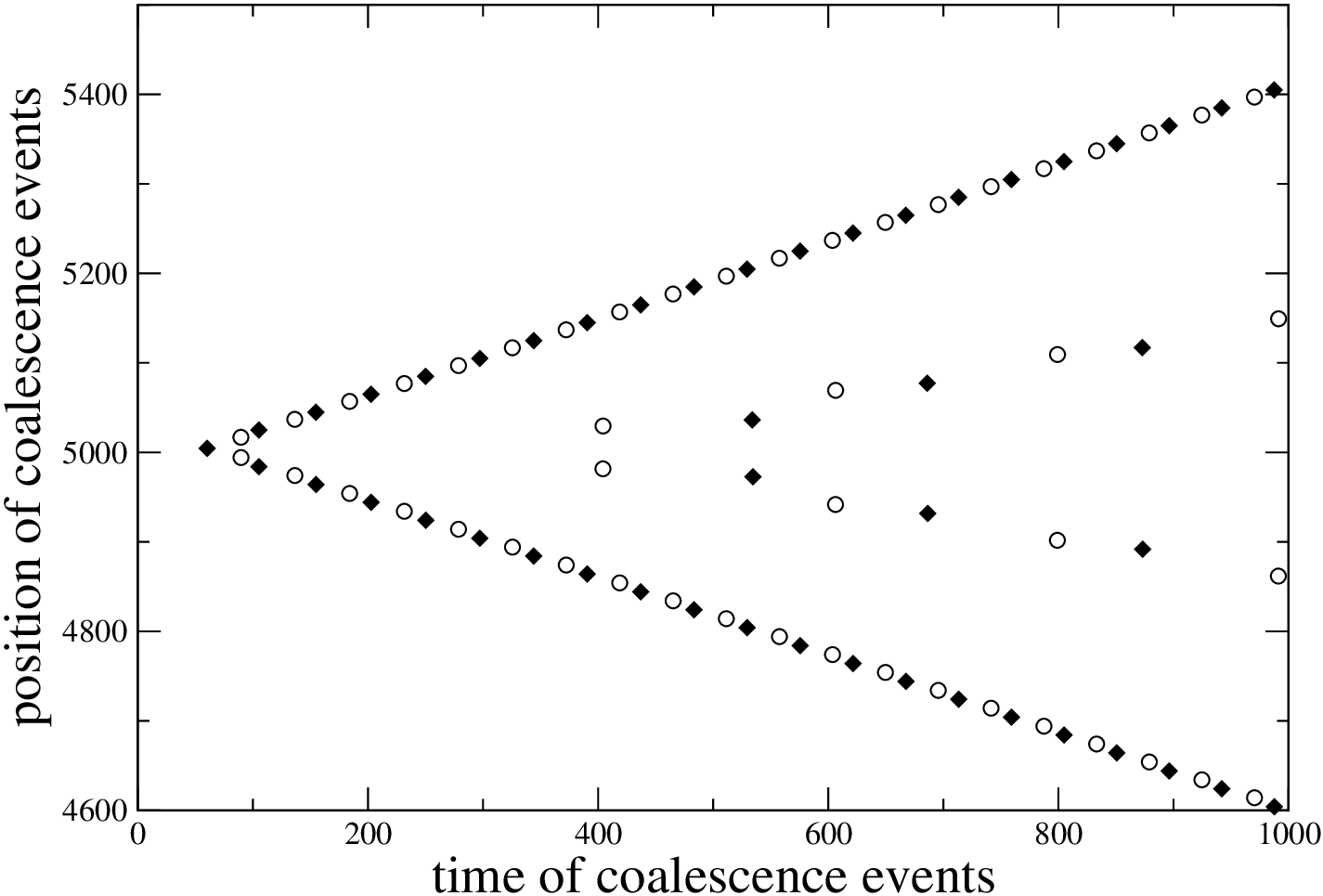}
\caption{
Temporal and spatial location of coalescence events following 
a single perturbation located in $x=5000$. The two sets of data (empty
circles and full diamonds) refer to perturbations of different
intensity and sign.
}
\label{fig_waves}
\end{figure}

\begin{figure}
\includegraphics*[width=12cm]{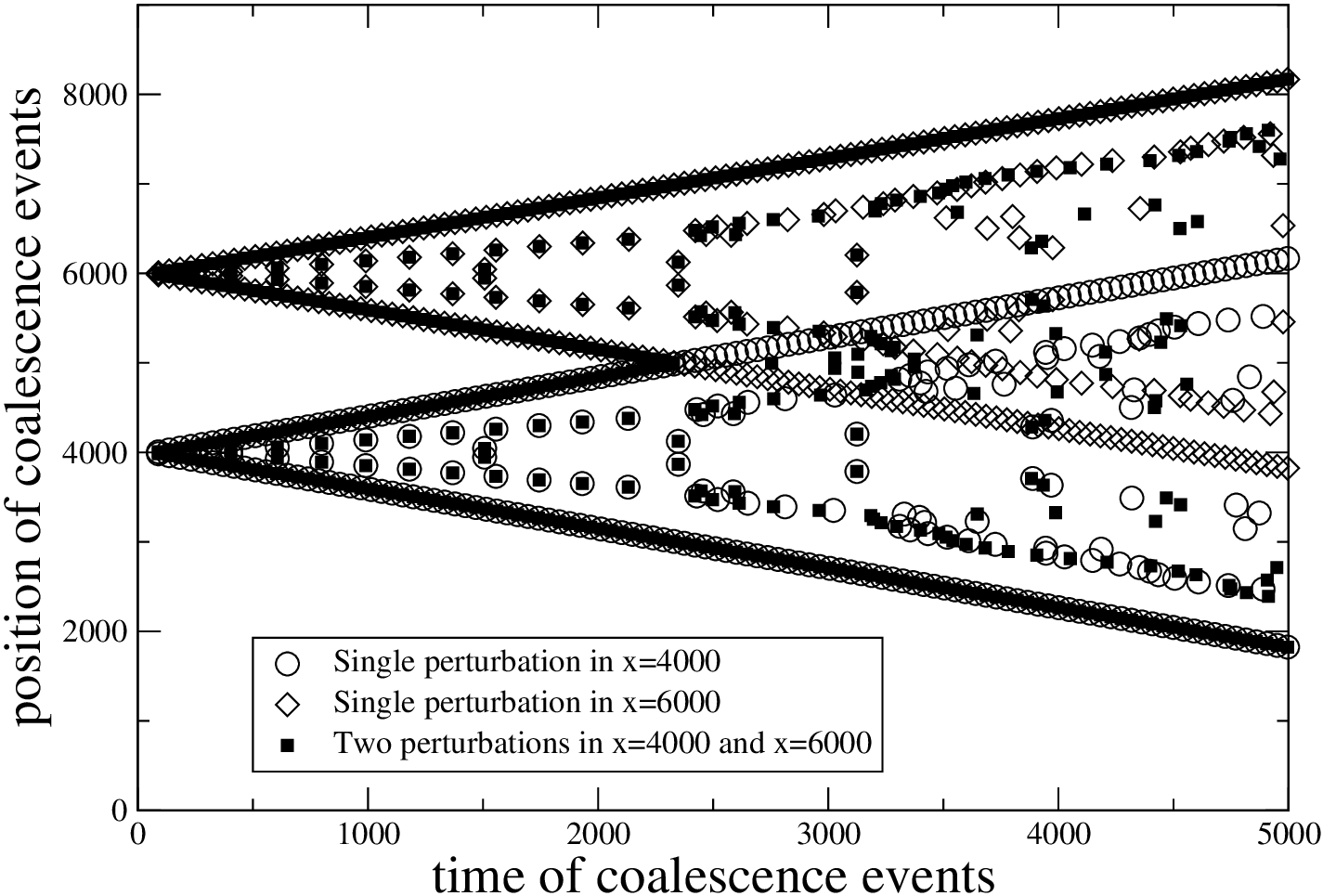}
\caption{
Full squares: Temporal and spatial location of coalescence events following
two simultaneous perturbations in $x=4000$ and $x=6000$.
These results are compared to the separate application of each
perturbation (empty symbols).
}
\label{fig_two_events}
\end{figure}

Consider a uniform distribution of $N$ particles, perturbed at a single
point: $\ell_i =\bar\ell$ for $i\ne N/2$ and
$\ell_{N/2} = \bar\ell + \Delta$.  As the system evolves, according to Eq.~(\ref{eq_l}),
the perturbation propagates on both sides of its origin, leaving behind a trail
of coalescence events.  These coalescence events take place at roughly equally spaced 
locations and time intervals, defining an apparent ``front", see Fig.~\ref{fig_waves}.
The speed $v$ of the front (the slope of the straight line in the figure) is independent
of the sign or magnitude of the perturbation, but seems to depend only on
the density of background particles, $v=c_1/\bar\ell$, with 
$c_1\approx 4$.
The frequency, or inverse time between two consecutive
coalescence events is roughly $c_2/\bar\ell^2$, with $c_2\approx 2$.
Therefore the coalescence front leaves  a diluted system behind, with
density $n\to n'=(1-c_2/c_1)n$, or about one half the original density.
Following these events, a second front of coalescence events sweeps through,
this time at roughly half the previous speed, due to the reduced density.
This second front, clearly visible in Fig.~\ref{fig_waves}, might be followed
yet by others, but their quality deteriorates fast as the background of
remaining particles distorts away from the original homogeneous spread.

Next, we ponder how two propagating fronts, originating from two distant
perturbations, interact.  In Fig.~\ref{fig_two_events} we plot simulations results
for this scenario.
One can see that coalescence fronts propagating in opposite directions
annihilate.  We have confirmed that annihilation takes place even when
the two perturbations are started at different times.  The rules for front
propagation and interaction seem very simple, and give us hope that they
might prove useful in shedding light on the kinetics of initially disordered particle
systems.

\section{Summary}

The initial motivation of our work was to study the so called
Conserved Kuramoto-Sivashinsky equation, Eq.~(\ref{cks}),
whose stability linear spectrum is $\omega=q^2-q^4$ and whose
numerical integration~\cite{FV} gives a coarsening process
with an exponent $n\simeq \fra{1}{2}$. The analysis of steady states
(Sec.~\ref{sec_ss})
tells us that stationary configurations have the form
$u_d(x)=u(x)+a$, where $u(x)$ is a $\lambda$-periodic function 
satisfying the differential equation
$u_{xx}=-u_x^2 -u$ and $a=a(\lambda)\simeq -\lambda^2/48$ 
is a constant fixed by the condition
$\langle u_d(x)\rangle=0$.

Numerics and theoretical background suggest that the function
$u(x,t)$ evolving according to the CKS eq. keeps close to steady states.
More precisely (Fig.~\ref{fig_arcs}), 
$u(x,t)$ appears to be similar to a continuous
piecewise function, where each {\it piece} is a portion of the universal
parabola $y(x)=a -(x-\bar x)^2/4$. Actually, connecting points
are vanishing regions of diverging positive curvature in the surface model. 
We have shown (Sec.~\ref{sec_pm}) 
that these angular points can be thought of
 as effective particles
and we derived the equations governing their dynamics, 
Eqs.~(\ref{eq_x},\ref{eq_l}).

Simulating the particles model is definitely easier than simulating 
the interface model and we have obtained (Sec.~\ref{sec_sim}) 
the coarsening law and the distribution
of interparticle distances. 
We also expect that future analytical treatment would sooner 
be addressed to the particles model and we suggest two main directions:
first, using the Fokker-Plank equation beyond the uncorrelated
intervals approximation, used in Sec.~\ref{sec_FP}; 
second, using the coalescence waves method introduced in Sec.~\ref{sec_cw}.

\ack
PP greatly acknowledges several useful discussions with his colleague 
Ruggero Vaia. The authors also thank Thomas Frisch and Alberto Verga for 
useful exchange of mails. Partial funding from the NSF (DbA) is gratefully
acknowledged.

\end{document}